\documentclass[aps,pra,a4paper,nofootinbib,preprint]{revtex4-1}

\makeatletter
\newcommand\footnoteref[1]{\protected@xdef\@thefnmark{\ref{#1}}\@footnotemark}
\makeatother

\usepackage{amsfonts,amsmath,amssymb,bm,tensor}
\usepackage{comment}
\usepackage{ifpdf}
\usepackage{slashed}
\usepackage{braket}
\usepackage[mathscr]{eucal}
\usepackage[utf8]{inputenc}
\usepackage{cancel}

\ifpdf
\usepackage{graphicx}     
\usepackage[bookmarksopen,colorlinks=true,linkcolor=bblue,citecolor=bblue,urlcolor=ppink]{hyperref}
\else     
\usepackage[dvipdfmx]{graphicx}     
\usepackage[dvipdfmx,bookmarksopen,colorlinks=true,linkcolor=bblue,citecolor=ppink,urlcolor=darkred]{hyperref}
\fi

\usepackage{color}
\definecolor{darkred}{rgb}{0.6,0,0}
\definecolor{darkgreen}{rgb}{0.992447,0.623778,0.034597}
\definecolor{ppink}{rgb}{1,0.4,0.4}
\definecolor{bblue}{rgb}{0.284602,0.317763,0.963947}
\definecolor{mygreen}{rgb}{0,0.7,0}
\definecolor{myred}{rgb}{1,0.3,0.4}
\definecolor{myblue}{rgb}{0.2,0.3,1}

	\newcommand{\Mpc}{\mathrm{Mpc}}
	\newcommand{\Gpc}{\mathrm{Gpc}}

	\newcommand{\MeV}{\mathrm{MeV}}

\newcommand{\bs}{\boldsymbol}
\newcommand{\tx}{\text}

\newcommand{\nn}{\nonumber\\}
\newcommand{\qcq}{\quad,\quad}

\newcommand{\df}{\text{d}}
\newcommand{\p}{\partial}

\newcommand{\msol}{{M_{\odot}}}

\begin{document}


\title{Effect of nonlinearity between density and curvature perturbations on the primordial black hole formation}

\author{Masahiro Kawasaki}
\affiliation{ICRR, The University of Tokyo, Kashiwa, 277-8582, Japan}
\affiliation{Kavli IPMU (WPI), UTIAS, The University of Tokyo, Kashiwa, 277-8583, Japan}
\author{Hiromasa Nakatsuka}
\affiliation{ICRR, The University of Tokyo, Kashiwa, 277-8582, Japan}
\affiliation{Kavli IPMU (WPI), UTIAS, The University of Tokyo, Kashiwa, 277-8583, Japan}

\begin{abstract}
We study the effect of the nonlinear relation between density and curvature perturbations on the formation of PBHs.
By calculating the variance and skewness of the density perturbation we derive the non-Gaussian property.
As a criterion for PBH formation, the compaction function is used and it is found that larger curvature perturbations are required due to the nonlinear effect.
We estimate the PBH abundance based on the Press–Schechter formalism with non-Gaussian probability density function during Radiation dominated era.
It is found that the nonlinear effect slightly suppresses the PBH formation and the suppression is comparable to that expected if the primordial curvature perturbation would have the local form of non-Gaussianity with nonlinear parameter $f_\tx{NL}\sim -1$.

\end{abstract}

\preprint{IPMU19-0029}
\date{\today}
\maketitle
\tableofcontents

\section{Introduction }
\label{sec:intro}

Primordial black holes (PBHs) has been attracting much interest since the LIGO-VIRGO collaboration succeeded in direct detection of the gravitational waves generated by mergers of binary black holes with masses around 30 $\msol$. 
Up to now about ten merger events have been found, from which the event rate is estimated as $53.2^{+58.5}_{-28.8}\Gpc^{-3}\tx{yr}^{-1}$~\cite{LIGOScientific:2018jsj}.
It is pointed out that PBHs with the abundance of $\Omega_\tx{PBH}/\Omega_\tx{DM}\sim 10^{-3}$ can be a promissing candidate for the merger events~\cite{Bird:2016dcv,Clesse:2016vqa,Sasaki:2016jop}.
PBHs are also an interesting candidate for the dark matter (DM). 
Although the available mass range for PBH to be DM is stringently constrained by the observations of $\gamma$-ray background, micro-lensing, CMB and others~\cite{Carr:2009jm,Allsman:2000kg,Tisserand:2006zx,Wyrzykowski:2011tr,Niikura:2017zjd,Ali-Haimoud:2016mbv,Graham:2015apa}, PBHs with  mass $M_\tx{PBH}\sim 10^{-13}\msol (\simeq10^{20}\tx{g})$ and $10^{-15}\msol (\simeq10^{18}\tx{g})$ can account for all DM of the universe becasue they avoid the micro(femto)-lensing constraint due to the wave effect and finite source size effect~\cite{Katz:2018zrn,Niikura:2017zjd}.

PBHs are also a unique tool to study the small scale perturbations and their generation in the early universe.
The formation of PBHs requires large density fluctuations of $\mathcal{O}(0.1)$ compared with $\mathcal{O}(10^{-5})$ at CMB scales, which is hardly realized in simple single-field inflation models and forces one to consider some mechanism to enhance small-scale density fluctuations such as hybrid inflation~\cite{Bugaev:2011wy}, double inflation~\cite{Inomata:2017vxo} and curvaton models~\cite{Kohri:2012yw,Ando:2017veq}.
PBHs are formed when overdense regions enter the horizon and gravitationally collapse~\cite{Carr:1975qj}.
The criterion of over-density required for PBH production was studied by the numerical relativistic simulations~\cite{Shibata:1999zs,Polnarev:2006aa}.
These simulations derived the threshold values on the curvature perturbation or the density perturbation using different gauge conditions. 
Harada {\it et. al.}~\cite{Harada:2015yda} compared those simulations and showed equivalence between their results.
In~\cite{Young:2014ana,Harada:2015yda} it was also pointed out that the threshold value of the curvature perturbation depends on environmental effects i.e. density profiles in the surrounding regions, which suggests that use of the averaged density perturbation is more suitable for the criterion than the curvature perturbation.

From point of view of model building, what we usually predict is the curvature perturbation power spectrum $\mathcal P_\zeta(k)$. 
To obtain the PBH abundance, one needs to evaluate the abundance of the overdense regions from the power spectrum.
The Press–Schechter formalism~\cite{Press:1973iz} is mostly used to count the number of overdense regions.
Other estimation based on the peak theory~\cite{Bardeen:1985tr} is also studied~\cite{Green:2004wb,Yoo:2018kvb}.
In the Press-Schechter formalism, we calculate variance $\sigma^2$ of the coarse-grained (averaged) density perturbation using a linear relation between the curvature and density perturbations, and then estimate the abundance of the overdense regions assuming that the coarse-grained density perturbation follows the Gaussian distribution with variance $\sigma^2$.
However, the relation between the curvature and density perturbations is non-linear, which could produce non-Gaussianity in the density perturbations. 

Although the non-Gaussianity is suppressed by the slow-roll parameters in the standard cosmology with the single field canonical inflation~\cite{Maldacena:2002vr}, some models like curvaton scenarios generate the perturbations with large non-Gaussianity. 
Since the non-Gaussianity changes the probability distribution of the perturbation, it can help or suppress the PBH formation.
Thus, there exist extensive researches considering the effect of  non-Gaussianity on the PBH formation~\cite{PinaAvelino:2005rm,Byrnes:2012yx,Young:2013oia,Sherkatghanad:2015rga,Young:2015cyn,Franciolini:2018vbk}.

In this paper, we revisit the relation between the curvature perturbation power spectrum and the PBH formation.
In particular, we study the effects of the nonlinear relation between the curvature perturbation $\zeta$ and density perturbation $\delta$ on the PBH formation.
Although the most previous calculations used the linear relation $\delta \propto \Delta \zeta$, the linear relation is not valid for large curvature perturbations.
The nonlinear relation can be obtained using the gradient expansion [see Eq.~(\ref{eq:delLWE})].
Thus, taking the nonlinearity into account, we estimate the threshold value on the average density perturbation from the result of the numerical simulations.
Here as the criterion for the PBH formation, we use the Compaction function $\mathcal{C}(r)$ introduced in Ref.~\cite{Shibata:1999zs}.
We also calculate the variance and skewness of the density perturbations and estimate the non-Gaussianity.
We thus obtain the PBH abundance based on the Press-Schechter formalism.

In the next section~\ref{sec:review}, we review the criterion of the PBH formation and estimate the formation rate using the Press–Schechter formalism and the linear density perturbation..
In Sec.~\ref{sec:NL}, the nonlinear effect on the PBH formation rate is analytically calculated.
We use the spherical harmonics expansion of the curvature perturbation and evaluate the variance and skewness of averaged density perturbation.
In Sec.~\ref{sec:result}, we numerically estimate the nonlinear effect for a simple power spectrum.
Sec~\ref{sec:cncl} is conclusion of this paper.

\section{PBH formation criterion and Press-Schechter formalism }
\label{sec:review}

When an overdensity region enters the horizon, a PBH is formed if the density perturbation grows gravitationally and overcomes the pressure.
One can evaluate the threshold value of the overdensity for the PBH formation by performing numerical relativity simulations~\cite{Shibata:1999zs,Polnarev:2006aa}.
In this section, we review the PBH formation criterion obtained by the results of the numerical relativity simulations following Ref.~\cite{Harada:2015yda}.

In order to calculate the superhorizon dynamics of the metric and energy-momentum tensors beyond the linear approximation, we use the long wavelength approximation (the gradient expansion), where the Einstein equations are expanded by the space derivative.
We introduce $\epsilon$ by $\p_i\to \epsilon \p_i$ and expand the equations in small $\epsilon$ on the flat background FLRW space. 
Since the superhorizon perturbations mainly contribute to the PBH formation, the gradient expansion is appropriate to the study of the formation criterion. 

The well-known criterion of the PBH formation is that the averaged density fluctuation over the horizon exceeds a certain threshold value.
Let us derived the threshold value following~\cite{Shibata:1999zs,Harada:2013epa,Harada:2015yda,Yoo:2018kvb}.
We consider the spherically symmetric curved space-time in the conformally flat spatial coordinates.
The metric  with the polar coordinate $(r,\theta, \varphi)$ is written as
\begin{align}
	\df^2 s=-(\alpha^2-\psi^2 a^2 \beta^2 r^2 )\df t^2
	+2\psi^4 a^2 \beta r\df r \df t
	+\psi^4 a^2 (\df r^2+r^2\df \Omega^2),
\end{align}
where $a(t)$ is the scale factor, $\df \Omega^2=\df \theta^2+\sin^2\theta d\varphi^2$, $\alpha$ and $\beta$ are the lapse function and the radial component of the shift vector.
In the comoving gauge, the comoving curvature perturbation $\zeta$ is given by $\psi=\exp(\zeta/2)$.\footnote{
In our definition, the density perturbation $\delta$ and the curvature perturbation $\zeta$ has the same sign at linear order in momentum space.}
The areal radius is given by $R(r)=\psi(r)^2a r=e^{\zeta(r)}a r$. 
Shibata and Sasaki~\cite{Shibata:1999zs} introduced the compaction function $\mathcal{C}(r)$ defined by
\begin{align}
	\label{eq:MSmass}
	\mathcal{C}(r)
	&=G\frac{\delta M(r)}{R(r)} 
\qcq
	\delta M(r)
	=
	4\pi \rho_b \int^r_0 \df x  R^2(x)R'(x) \delta(x),
\end{align}
where $\rho_b$ is the energy density of the background universe, $\delta$ is the density perturbation ($\delta=(\rho-\rho_b)/\rho_b$) and $'$ denotes $d/dr$.
In this paper, the subscript "$b$" denotes the background value. 
In the long-wave approximation the relation between the comoving density perturbation $\delta$ and the curvature perturbation $\zeta$ is written as~\cite{Harada:2015yda}
\begin{align}\label{eq:delLWE}
	\delta(r)=
	-\frac{4(1+w)}{5+3w}
	  \frac{\Delta (e^{\zeta/2}) }{e^{5\zeta/2}}\left(\frac{1}{aH_b} \right)^2
		+\mathcal{O}(\epsilon^3)
\end{align}
where $w=P/\rho$ is the equation of state and $H_b$ is the background Hubble parameter. 
In this paper, we focus on the radiation dominated era ($w=1/3$).

Since $R\propto a$ and $\delta M\propto \rho_b R (aH_b)^{-2}\propto a$, $\mathcal{C}(r)$ is independent of $a$ and $H_b$. 
$\mathcal{C}(r)$ can be calculated using the initial profile of $\zeta(r)$ since it is frozen in the super horizon.
For a given profile, $\mathcal{C}(r)$ has the maximum value $\mathcal{C}_m$ at some  point $r=r_m$. 
In the constant mean curvature slicing, the numerical simulation gives the threshold value  of the PBH formation as $\mathcal{C}_{m;\tx{CMC}}^{th}\simeq 0.4$ during the radiation dominated eta \cite{Shibata:1999zs}.
Once the initial profile gives $\mathcal{C}_m$ larger than the threshold value, we assume that the region collapses into the PBH.
Some uncertainty on the threshold value comes from the different shapes of the initial profiles~\cite{Shibata:1999zs}.
Detailed analysis about the shape of the profile was performed in~\cite{Harada:2015yda,Nakama:2013ica} and it was shown that the sharp profiles tend to require large overdensities to collapse. In this paper, we only consider the condition on the averaged density, for simplicity.
In the comoving gauge, the threshold value is given by
$\mathcal{C}_{m;\tx{Com}}^{th}=\frac{2}{3}\mathcal{C}_{m;\tx{CMC}}^{th}\simeq 0.27$
\cite{Harada:2015yda}.

The averaged density fluctuation $\bar \delta$ is defined as
\begin{equation}
    \bar{\delta} = \frac{4\pi \int_{0}^{r}dx R(x)^2 R'(x)\delta(x)}
    {4\pi R(r)^3/3},
    \label{eq:bardeldef}
\end{equation}
which is related to the compaction function $\mathcal{C}(r)$ as~\cite{Harada:2015yda}
\begin{align}\label{eq:Cdel}
	\mathcal{C}(r)
	=&\frac{1}{2}\bar \delta(r) (H_b R(r))^2.
\end{align}
Since the density perturbation depends on the cosmic time $t$, we should specify the time to evaluate $\bar \delta$. 
The usual choice is the horizon crossing time $H_bR(r_m)=H_b a r_me^{\zeta(r_m)}=1$, which depends on the curvature profile itself.
At the horizon crossing time, the threshold value of the density perturbation is given by $ \bar \delta^{th}=2\mathcal{C}^{th}_{m;\tx{Com}}\simeq0.53$ which is adopted in our paper.

Now we estimate the PBH formation rate in the linear order calculation based on the Press–Schechter formalism~\cite{Press:1973iz}. 
The density perturbation is related to $\zeta$ in momentum space as
\begin{align}
	\label{eq:lindel}
	\delta_k=	\tfrac{4}{9}	\left(\tfrac{k}{aH_b}\right)^2\zeta_k +\mathcal O(\zeta^2).
\end{align}
The averaged density fluctuation over the radius $\rho$ is given by
\begin{align}
	\label{eq:deltaW}
	\bar \delta(\bs x,\rho)=
	\int\df^3 y 
	W(|\bs x-\bs y|;\rho)
	\delta(\bs y)
	=\int \frac{\df^3 k}{(2\pi)^3}
	\tilde W(k \rho)e^{i\bs k\cdot \bs x}\delta_k
\end{align}
where $W(r;\rho)$ and $\tilde W(k \rho)$ is the window functions in the real and momentum space.
It is pointed out that the choice of the window function leads to large uncertainties~\cite{Ando:2018qdb}.
When we use the average density fluctuation defined by Eq.~(\ref{eq:bardeldef}), we should use the real space top-hat window,
\begin{align}
    W(x;\rho) & = \left(\frac{4\pi \rho^3}{3}\right)^{-1}\theta(\rho-x), \\
    \tilde{W}(k\rho) &= 3\left(\frac{\sin(k\rho)-k\rho\cos(k\rho)}{(k\rho)^3}\right). 
\end{align}
In the subhorizon, the perturbations start to oscillate and their amplitude decreases.
The time evolution of the perturbations is described by the transfer function given by
 \begin{align}
	 T(k\eta)=&
	 3\frac{\sin(k\eta/ \sqrt{3})-(k\eta/ \sqrt{3})\cos(k\eta/\sqrt{3})}{ (k\eta/\sqrt{3})^3 } ,
 \end{align}
 which is used together with the real space top hat window function because high momentum modes give a significant contribution to $\bar\delta(\mathbf{x},\rho)$.\footnote{
For the Gaussian window function, which is often used in the literature, high momentum modes are exponentially suppressed and hence the effect of the transfer function is negligible.}
Assuming that $\zeta$ is Gaussian variable, the probability density function of $\bar \delta$ is given by
\begin{align}
	P_{k}(\bar \delta)
	&=\frac{1}{\sqrt{2\pi\sigma_k^2}}\exp\left( -\frac{\bar\delta^2 }{2\sigma_k^2 } \right)
	,
	\label{eq:gaussianP}
\\
	\sigma_k^2&=
	\braket{\bar \delta(x,\rho=k^{-1})^2}
	\simeq
	\int \df (\ln p)
	\frac{16}{81}
	|\tilde W(p/k)|^2
	\left(\frac{p}{aH_b}\right)^4
	\left|T\left(\tfrac{p}{aH_b}\right)\right|^2
	\mathcal P_\zeta(p)
	\label{eq:oldsigma}
\end{align}
In the linear order approximation, the horizon crossing condition is given by $aH_b=k$, which is slightly different from the nonlinear one. 
The PBH formation rate is given by
\begin{align}\label{eq:PSform}
	\beta(k)
	&=\int^\infty_{\bar \delta^{th}}	\df \bar \delta~
	P_k(\bar \delta)
\end{align}

The mass of PBHs is nearly equal to the horizon mass when the PBHs are formed~\cite{Carr:1975qj}.
Thus, in the radiation dominated era, the PBH mass is written as 
\begin{align}\label{eq:Mkrel}
	M
	&=
	\gamma \rho_{r}(t_\tx{HC}) \frac{4\pi H^{-3}}{3}|_{k=aH}
	\nn&\simeq
	30 \msol
	\frac{\gamma}{0.2}
	\left( \frac{g_*}{10.75} \right) ^{-1/6}
	\left( \frac{k}{3.43\times 10^{5}\,\Mpc^{-1}} \right)^{-2}
	\nn&\simeq
	30 \msol 
	\frac{\gamma}{0.2}
	\left( \frac{g_*}{10.75} \right)^{-1/2}
	\left( \frac{T}{31.6\,\MeV } \right)^{-2}.
\end{align}
where $g_*$ is relativistic degree of freedom, $\rho_{b}(t_\tx{HC})$ is the background radiation energy density at horizon crossing and $\gamma$ is the ratio of PBH mass to the horizon mass.
In this paper, we take the analytical estimation $\gamma=3^{-3/2}\simeq0.2$ \cite{Carr:1975qj}. 
The mass spectrum is then given by
\begin{align}
	\label{eq:massspec}
	f(M)
	&=\frac{\df \Omega_\tx{PBH} }{\df \ln M} 
	\frac{1}{\Omega_\tx{DM}}
	=
	\frac{\beta(k(M))}{1.8 \times 10^{-8}}
	\left( \frac{\gamma}{0.2} \right)^{3/2}
	\left( \frac{10.75}{g_{*;30\MeV}} \right)^{1/4}
	\left( \frac{0.12}{\Omega_\tx{DM}h^2} \right)
	\left( \frac{M}{ \msol } \right)^{-1/2}
	.
\end{align} 

\section{Effect of nonlinearity} 
\label{sec:NL}

\begin{figure}[t]
    \centering 
    \includegraphics[width=.60\textwidth]{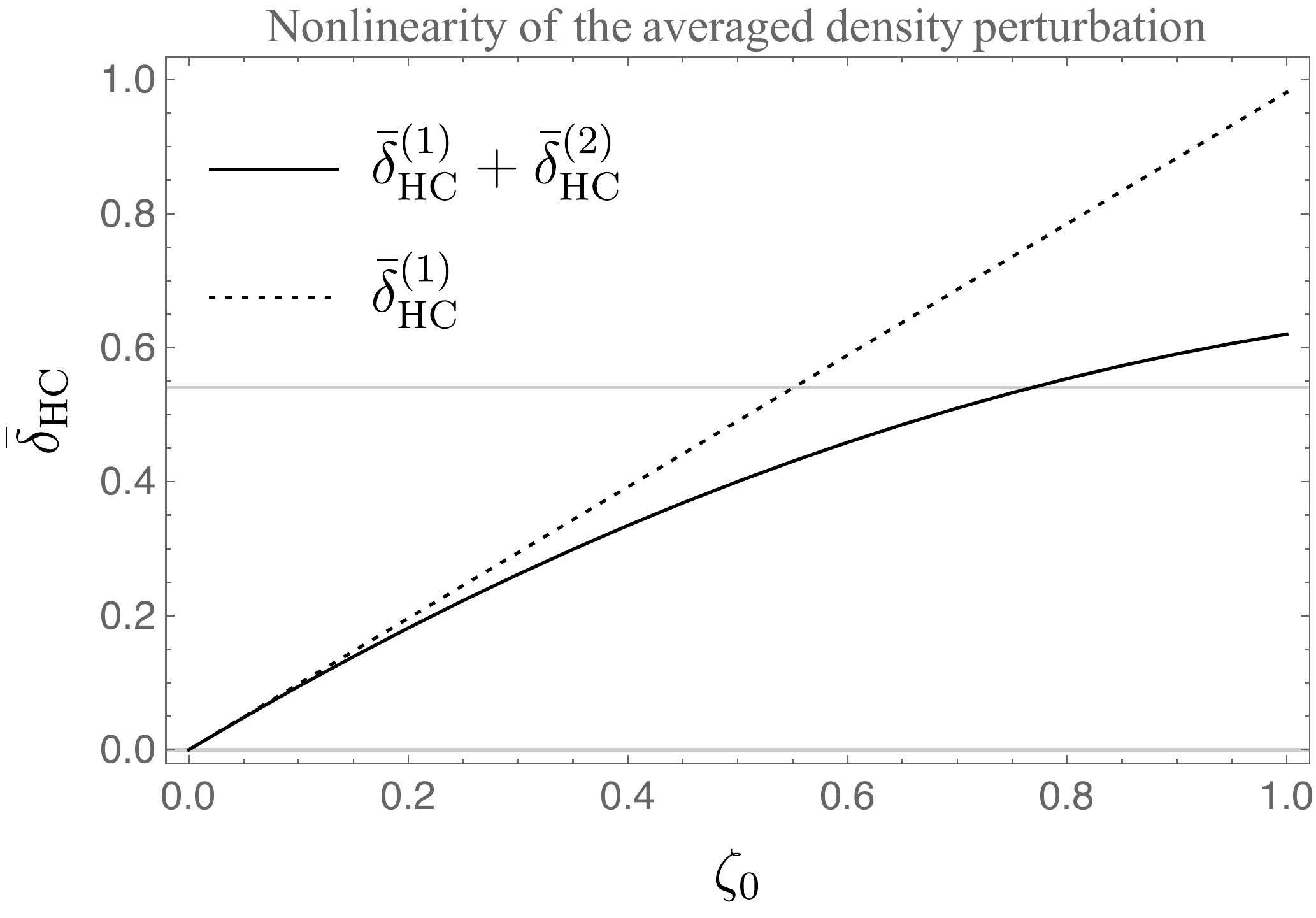}
    \caption{Nonlinearity of the averaged density perturbation in Eq.\eqref{eq:delHC}.
    We use the test profile $\zeta (r)=\zeta_0 e^{-r^2/2}$
    and plot $\bar{\delta}_\tx{HC}( r_m=\sqrt{2} )$ for each $\zeta_0$.
    The horizontal line is threshold value $\bar \delta^{th}=0.53$.
	}
	\label{fig:delform}
\end{figure}

Now we consider the effect of the nonliear relation between $\bar \delta$ (or $\mathcal{C}$) and $\zeta$ found in  Eqs.~\eqref{eq:delLWE}, \eqref{eq:bardeldef} and \eqref{eq:Cdel}.
In the Press–Schechter formalism, we evaluate the probability density function of the averaged density for a given horizon scale $r_m$.
When the averaged density exceeds $\bar \delta^{th}$, the region collapses into a PBH.\footnote{
Although $r_m$ is defined for each initial profile in numerical simulations, we now take $r_m$ as horizon scale to perform the statistical estimation.
This picture has the double counting problem when $\mathcal{C}(r)$ has many maximum points.
The smooth window function or the transfer function helps to avoid this problem by suppressing the subhorizon modes.}
At horizon crossing $H_bR(r_m)=1$, the averaged density perturbation is calculated as~\cite{Harada:2015yda}
\begin{align}
\label{eq:delHC}
	\bar \delta _{\tx{HC}}(r_m) 
	=&~
	\left(	\frac{4\pi R(r_m) ^3}{3}	\right) ^{-1}
	\left(
	4\pi  \int^{r_m}_0 \df r R(r)^2  R'(r) \frac{-8}{9}	\frac{\Delta \psi(r)}{\psi(r)} \frac{r^2_m\psi(r_m) ^4 }{\psi(r)^4}
	\right)	
	+\mathcal{O}(\epsilon^3)
\\	=&~
	\left( \frac{4\pi r_m^3}{3} \right)^{-1}
	4\pi \int ^{r_m}_0\df r r^2 r_m^2 
	\left(-\frac{4}{9}\right)   \nonumber
\\  
    &\times\left[
	\Delta\zeta(r)+
	(\zeta(r)-\zeta(r_m))\Delta\zeta(r)
	+\frac{1}{2}\zeta'(r) (5\zeta'(r)+2 r\zeta''(r) )
	\right]
	+\mathcal{O}(\epsilon^3)
\\ =&~ \frac{2}{3}[1-(1+r_m\zeta'(r_m))^2]
    +\mathcal{O}(\epsilon^3)
    \label{eq:delintegrated}
\\=&~
	\bar \delta _{\tx{HC}}^{(1)}(r_m)
	+\bar \delta _{\tx{HC}}^{(2)}(r_m) 
	+\mathcal{O}(\epsilon^3).
\end{align}
where $\bar \delta _{\tx{HC}}^{(n)}$ denotes the $\mathcal O(\zeta^n)$ term.
Thus, the averaged density is a nonlinear function of $\zeta$.
When we substitute the threshold value $\bar\delta^{th}=0.53$, the linear order solution is given by $r_m\zeta'(r_m)\simeq -0.40$ and the exact one is $r_m\zeta'(r_m)\simeq -0.55$, which are differ by a factor about $1.4$.
Fig.\ref{fig:delform} shows the contribution of $\bar \delta _{\tx{HC}}^{(1)}(r_m) $ and $\bar \delta _{\tx{HC}}^{(2)}(r_m) $ for $\zeta (r)=\zeta_0 e^{-r^2/2}$ at $r_m=\sqrt{2}$.
The non-linearity slightly suppresses $\bar \delta_\tx{HC}$ compared to the linear order calculation $\bar \delta _{\tx{HC}}^{(1)}(r_m)$.
This suggests that PBHs are difficult to form and hence the formation rate decreases due to non-linearity.
We will confirm it using a non-Gaussian distribution function.


In the following section, we treat $\zeta$ and $\bar \delta_\tx{HC}$ as stochastic variables and
estimate the probability density function of $\bar \delta_\tx{HC}$ to calculate the PBH formation rate Eq.\eqref{eq:PSform}.
Since the criterion Eq.\eqref{eq:delHC} is obtained assuming spherical symmetry, we use the spherical harmonics expansion~\cite{Yoo:2013tc} as\footnote{
Since PBH formation requires the highly curved space, the scale $k$ may depend on the curvature of each region. 
Here we simply assume that the Fourier transformation on the flat background space.
}
\begin{align}
	\zeta(\bs x)
	~=~&\int^\infty_0 \df k
	\sum_{lm} \sqrt{\frac{2}{\pi}} k j_l(kr)Y_{lm}(\hat {\bs x})
	\zeta_{lm}(k),
	\label{eq:zetamomentum}
	\\
	\zeta_{lm}(k)
	~=~&
	\int\df^2 \Omega_k Y^*_{lm}( {\bs k}/|\bs k|) \zeta(\bs k),
	\\
	\braket{\zeta_{l_1m_1}(k_1)\zeta_{l_2m_2}(k_2)^*}
	~=~&\delta_{l_1l_2}\delta_{m_1m_2}
	\delta(k_1-k_2)
	\frac{2\pi^2}{k_1^3}
	\mathcal P_\zeta(k_1),
\end{align}
where $j_l(kr)$ is the spherical Bessel function of the first kind, $Y_{lm}(\hat {\bs x})$ is the spherical harmonics and $\int \df \Omega_k$ is the integration over the direction of $\bs k$.
Due to spherically symmetry, only $\zeta_{00}(k)$ has a non zero value. 
In this paper, we assume that $\zeta_{00}(k)$ is a Gaussian variable, which is justified for slow-roll inflation~\cite{Maldacena:2002vr}.
On the other hand, since $\bar \delta _{\tx{HC}}(r_m) $ is the nonlinear function of $\zeta$, $\bar \delta_\tx{HC}$  follows the non-Gaussian distribution. 
Following the treatment of non-Gaussianity of the PBH formations \cite{Young:2015cyn,Franciolini:2018vbk}, we evaluate the probability density function using the variance $\sigma$ and the skewness $\mu_3$ defined by \cite{Ezquiaga:2018gbw}
\begin{align}
    \sigma(r_m)^2&= 
    \braket{\bar \delta_\tx{HC}(r_m)^2}-\braket{\bar \delta_\tx{HC}(r_m)}^2,
    \label{eq:variancedef}
\\
    \mu_3(r_m)
   &=(\sigma(r_m))^{-3}
    \left(
    \braket{\delta _{\tx{HC}}(r_m)^3}
    -3\braket{\delta _{\tx{HC}}(r_m)^2}
    \braket{\delta _{\tx{HC}}(r_m)}
    +2\braket{\delta _{\tx{HC}}(r_m)}^3
    \right).
    \label{eq:skewdef}
\end{align}

At first, we simplify $\bar \delta _{\tx{HC}}^{(1)}(r_m)$ and $\bar \delta _{\tx{HC}}^{(2)}(r_m)$ using Eq.\eqref{eq:zetamomentum} as
\begin{align}
	\bar \delta _{\tx{HC}}^{(1)}(r_m)
	&=
	\left(	\frac{4\pi  r_m^3}{3}	\right) ^{-1}
	\left(
	4\pi  \int^{ r_m}_0 \df r  r^2
	\frac{-4  }{9}
	r^2_m \Delta \zeta(r)
	\right)
\nonumber
\\&=
	\frac{4 }{9}
	\int\df\ln(r_mp) ~(r_mp)^4
	\tilde W(r_mp)
	\frac{r_m^{-2}\zeta_{00}(p)}{\sqrt{2}\pi},
\\
	\bar \delta_\tx{HC}^{(2)}(r_m)	
	&= 
	\left( \frac{4\pi r_m^3}{3} \right)^{-1}
	4\pi \int ^{r_m}_0\df r r^2 r_m^2 
	\times
	\frac{-4}{9}
	\left\{
	\Delta\zeta(\zeta-\zeta(r_m))
	+\frac{1}{2}\zeta' (5\zeta'+2 r\zeta'' )
	\right\} \nonumber
	\\
	&= 
	\frac{-4}{9}
	\int^\infty_0 \df (p_1r_m) \df (p_2r_m)~ 
	\frac{r_m^{-4}}{2\pi^2} \zeta_{00}(p_1)\zeta_{00}(p_2)\nonumber 
\\
	& \times \frac{3}{2 }
	\left[ \sin(p_1r_m)-(p_1r_m)\cos(p_1r_m) \right]
	\left[ \sin(p_2r_m)-(p_2r_m)\cos(p_2r_m) \right],
\end{align}
where $\tilde{W}(z)(=3(\sin z-z \cos z)/z^3)$ is the Fourier transformation of the real space top-hat window function. 
Although the real space top-hat window function is derived from the definition of the compaction function, it may have some troubles. 
(1) Our calculation is only valid in small $k$ because of the long wavelength approximation. 
On the other hand, the averaged density with top-hat window function tends to include a significant contribution from large $k$ modes compared to the Gaussian window function or the momentum space top-hat window function. 
(2) As pointed out in~\cite{Yoo:2018kvb} the contribution of large $k$ in the top-hat window function comes from the sharp edge in the real space.
Thus, we need to suppress the high momentum modes, for example, multiplying the transfer function or using the Gaussian or the momentum space top-hat window function instead of the real space top-hat window function.
In this paper, since we concentrate on the non-linearity of the compaction function, we continue to use the real space top-hat window function and multiply it by the transfer function, $\mathcal{P}_\zeta(p) \to |T(r_m p)|^2\mathcal{P}_\zeta(p)$ in an ad-hoc way.\footnote{
  Using monochromatic spectrum $\zeta(r)=A \sin(kr)/(kr)$, $\mathcal{C}(r)$ shows approximately periodic behavior and has many peaks.
  The transfer function helps to pick up only the first peak with the smallest $r_m$.
  Instead, we can use the step function to pick up the first peak, which results in the larger PBH abundance.
}

The variance Eq.\eqref{eq:variancedef} up to $\mathcal{O}(\zeta^2)$ is written as
\begin{align}
	\sigma(r_m)^2
	&=
	\braket{\,(\bar \delta_\tx{HC}^{(1)}(r_m))^2\,}+\mathcal O(\zeta^4)
\\& =
	\left( 	\frac{4}{9}	  \right)^2
	\int \df (\log r_m p)~
	|\tilde W(r_m p)|^2
	(r_m p)^4
	|T(r_m p)|^2\mathcal{P}_\zeta(p)
	+\mathcal O(\zeta^4) .
	\label{eq:del2}
\end{align}
Because we only include $\braket{\bar \delta_\tx{HC}^{(1)2}(r_m)}$, $\sigma^2$ reproduces the linear order calculation in Eq.\eqref{eq:oldsigma}.
As for the skewness we calculate $\mu_3(r_m)$ up to $\mathcal O(\zeta^4)$,
\begin{align}
    \mu_3(r_m)
	&=\sigma^{-3}
	\left[
	3\braket{(\bar \delta_\tx{HC}^{(1)}(r_m))^2\,
	\bar \delta_\tx{HC}^{(2)}(r_m)}
	-
	3
	\braket{(\bar \delta_\tx{HC}^{(1)}(r_m))^2}
	\braket{
	\bar \delta_\tx{HC}^{(2)}(r_m)}
	\right]
	+\mathcal O(\zeta^6) 
    \nonumber 
\\&=
    -9	\sigma^{-3}\left( \frac{4}{9}\right)^3
	\int\df (p_1r_m) \df (p_2r_m)
	\tilde W(p_1r_m)\tilde W(p_2r_m)
	|T( p_1r_m)|^2
	\mathcal P_\zeta(p_1) 
	|T(p_2r_m )|^2
	\mathcal P_\zeta(p_2) 
\nonumber
\\&\quad\times 
	\left( \sin(p_1r_m)-(p_1r_m)\cos(p_1r_m) \right)
		\left( \sin(p_2r_m)-(p_2r_m)\cos(p_2r_m) \right)
\\&=		
	-\frac{9}{4}\sigma(r_m)
	\label{eq:del3}
\end{align}
Since the linear and nonlinear terms both depend on $r_m \zeta'(r_m)$ in Eq.\eqref{eq:delintegrated}, $\mu_3$ is expressed by $\sigma$.
Using the variance $\sigma^2(r_m)$ and the skewness $\mu_3(r_m)$, we approximately obtain the probability density function of $\bar \delta_\tx{HC}(r_m)$.
When the skewness is regarded as the small perturbation, $\bar \delta_\tx{HC}$ is wriiten with use of the Gaussian variable $\delta_g$ as
\begin{align}
  \bar \delta_\tx{HC}[\delta_g]
  =\delta_g
  +\frac{
  \mu_3(r_m)
  }{6\sigma(r_m)}
  (\delta_g^2-\sigma^2)  
  ,
  \label{eq:delhcapprox}
\end{align}
where $\delta_g$ follows the Gaussian probability density function $P_g(\delta_g)$,
\begin{align}
    P_g(\delta_g)
    =\frac{1}{\sqrt{2\pi}\sigma(r_m) } 
    \exp\left(
    -\frac{1}{2} \frac{\delta_g^2}{\sigma(r_m)^2}
    \right).
\end{align}
Therefore, the probability density function of $\bar \delta_\tx{HC}$ is obtained as~\cite{Young:2013oia,Byrnes:2012yx,Young:2015cyn}
\begin{align}
	P_{NG}(\bar \delta_\tx{HC})
	&=\sum_{i=\pm}
	\left| \frac{\df \delta_{g;i}(\bar \delta_\tx{HC})}{\df \bar \delta_\tx{HC}}	\right|
	P_g(\delta_{g;i}(\bar \delta_\tx{HC}))
	,
\end{align}
where $\delta_{g;\pm}(\bar \delta_\tx{HC})$ are two solutions of $\bar \delta_\tx{HC}=\bar \delta_\tx{HC}[\delta_g]$ in Eq.\eqref{eq:delhcapprox} given by
\begin{align}
	\delta_{g;\pm}(\bar \delta_\tx{HC})
	=\frac{3\sigma}{\mu_3}
	\left(
	-1\pm \sqrt{
		1+\frac{2\mu_3}{3} 
		\left(  \frac{\mu_3}{6}  +  \frac{\bar \delta_\tx{HC}}{\sigma } \right)
	}
	\right).
\end{align}
We then rewrite the PBH formation rate Eq.\eqref{eq:PSform} as
\begin{align}
	\nonumber
	\beta(r_m^{-1})
	=&\int_{\bar \delta_\tx{HC}>\bar \delta^{th}} P_{NG}(\bar \delta_\tx{HC}) \text{d} \bar \delta_\tx{HC}
	=\int_{\bar \delta_\tx{HC}[\delta_g]>\bar \delta^{th}} P_{g}( \delta_{g} ) \text{d} \delta_{g}
\\[0.5em]
    \nonumber=&
	\begin{cases}
	\int_{ \delta_{g;+}(\bar \delta^{th}) }^{ +\infty } P_{g}( \delta_{g} ) \text{d} \delta_{g}
	+
	\int_{ -\infty}^{ \delta_{g;-}(\bar \delta^{th}) } P_{g}( \delta_{g} ) \text{d} \delta_{g}
	&~~~~~\text{for }\mu_3>0
	\\[0.4em]
	\int_{ \delta_{g;+}(\bar \delta^{th} ) }^{ \delta_{g;-}(\bar \delta^{th}) } P_{g}( \delta_{g} ) \text{d} \delta_{g}
	& ~~~~~\text{for }\mu_3<0
	\end{cases}
	\\	=&
	\begin{cases}
	1
	-\frac{
		\text{Erf}(\delta_{g;+}(\bar \delta^{th})/\sqrt{2\sigma^2})
		-\text{Erf}(\delta_{g;-}(\bar \delta^{th})/\sqrt{2\sigma^2})
	}{2}
	& ~~~~~\text{for }\mu_3>0
	\\[0.4em]	
	\frac{
		\text{Erf}(\delta_{g;-}(\bar \delta^{th})/\sqrt{2\sigma^2})
		-\text{Erf}(\delta_{g;+}(\bar \delta^{th})/\sqrt{2\sigma^2})
	}{2}
	& ~~~~~\text{for } \mu_3<0.
	\end{cases}
	\label{eq:probnongau3}
\end{align}
with a error function $\tx{Erf}(x)=\frac{2}{\sqrt{\pi}}\int^z_0 \df t ~e^{-t^2}$.

To calculate the mass spectrum, we need to relate the PBH mass and the horizon scale $r_m$. 
The PBH mass is more directly related to areal radius $R(r_m)$, not  $r_m$ because the horizon scale is given by $H_b^{-1}= R(r_m)$. 
The difference between two scales may change the PBH mass by $e^{2\zeta(r_m)}= \mathcal O(1)$.
However, other uncertainties also change masses of PBHs.
For example, the critical phenomena of PBH formation~\cite{Niemeyer:1999ak} and the choice of the initial profiles leads to uncertainties of $\mathcal{O}(1)$.
In this paper, thus, we obtain PBH mass using Eq.\eqref{eq:Mkrel} with $k_m=r_m^{-1}$ as an approximation.


\section{Numerical results}
\label{sec:result}

\begin{figure}[t]
	\centering 
	\includegraphics[width=.75\textwidth]{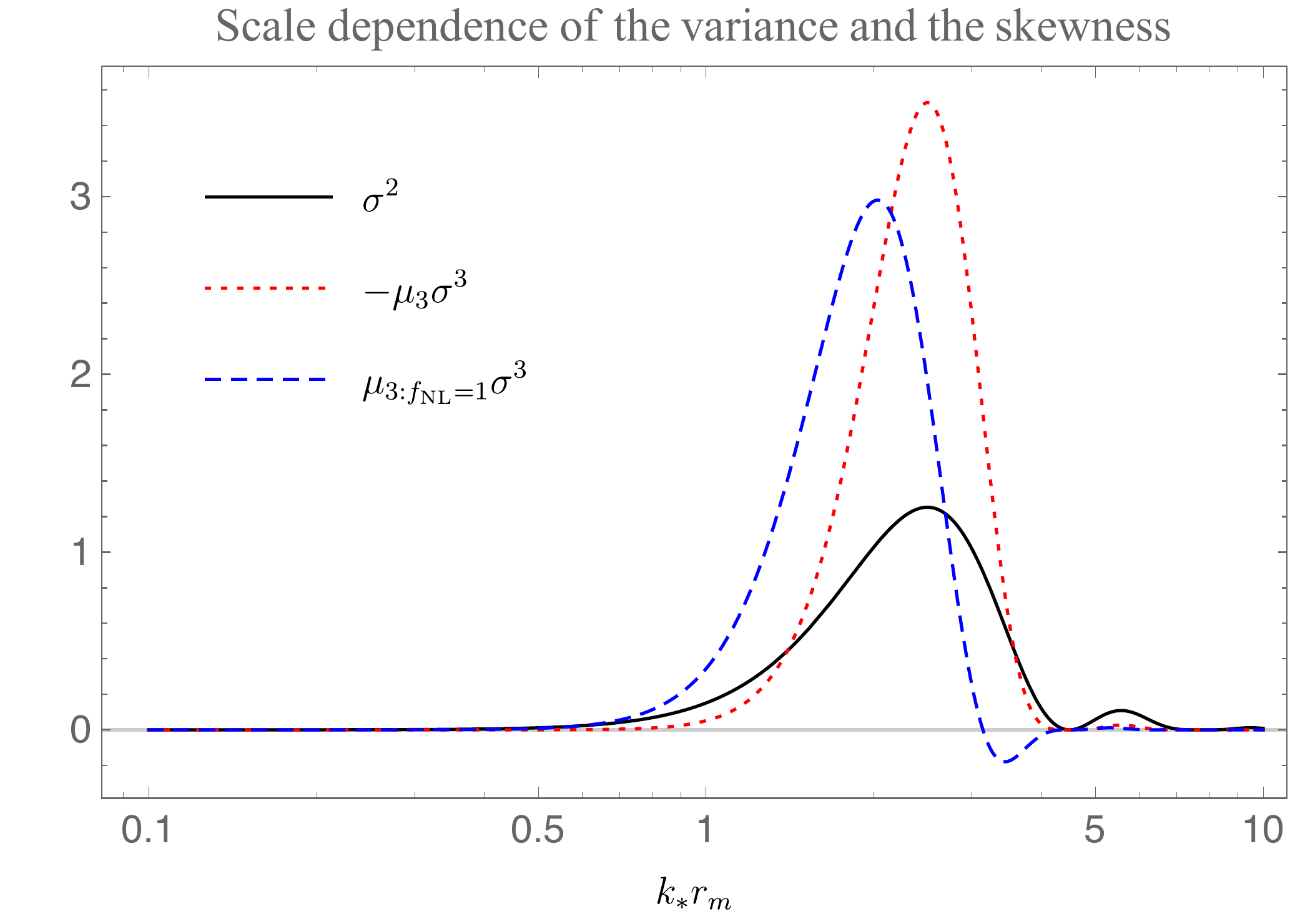}
	\caption{
	Scale dependence of the variance and the skewness at $A=1$.
	 We plot $\sigma^2$ in Eq.\eqref{eq:del2num} and  $\mu_3$ in Eq.\eqref{eq:del3num} for each scale $k_* r_m$. 
	We also plot $\mu_3$ derived from the primordial non-Gaussianity of $\zeta$ with local type bispectrum $f_\tx{NL}=1$ in Eq.\eqref{eq:del3fnlnum}.
	}
	\label{fig:cumulants}
\end{figure}

To estimate the nonlinear effect, we calculate PBH mass spectrum for the following simple curvature power spectrum:
\begin{align}\label{eq:powerspec}
	\mathcal P_\zeta(k)=Ak_*\delta(k-k_*).
\end{align}
Inserting $P_\zeta(k)$ into Eq.\eqref{eq:del2} and Eq.\eqref{eq:del3}, the variance and the skewness are written as
\begin{align}
	\label{eq:del2num}
	\sigma^2(r_m)
	&=
	\left( 	\frac{4}{9}	  \right)^2
	|\tilde W(r_m k_*)|^2|T(r_m k_*)|^2
	(r_m k_*)^4  A .
\\
	\label{eq:del3num}
	\mu_3(r_m)
	&=
	-9(\sigma(r_m))^{-3}
	\left( \frac{4}{9}\right)^3 
	|\tilde W(k_*r_m)|^2
	|T(r_m k_*)|^4
	(k_*r_m)^2 
\nonumber \\
	& ~~\times \left[ \sin(k_*r_m)-(k_*r_m)\cos(k_*r_m) \right]^2
	\, A^2.
\\&=-|\tilde W(r_m k_*)||T(r_m k_*)|
	(r_m k_*)^2  \sqrt{A} .
\end{align}
The negative $\mu_3$ suppresses the PBH formation rate, as expected.  
We show the scale dependence of the variance and the skewness with $A=1$ in Fig.\ref{fig:cumulants}.
The black solid line represents the variance of $\bar \delta_\tx{HC}$, which has the maximum value at $k_*r_m\sim 2.5$. 
The dotted red line shows $-\mu_3(r_m)$. 

\begin{figure}[t]
	\centering 
	
	\begin{tabular}{c}
		
		\begin{minipage}{0.4\hsize}	\begin{center}
			\includegraphics[width=.88\textwidth]{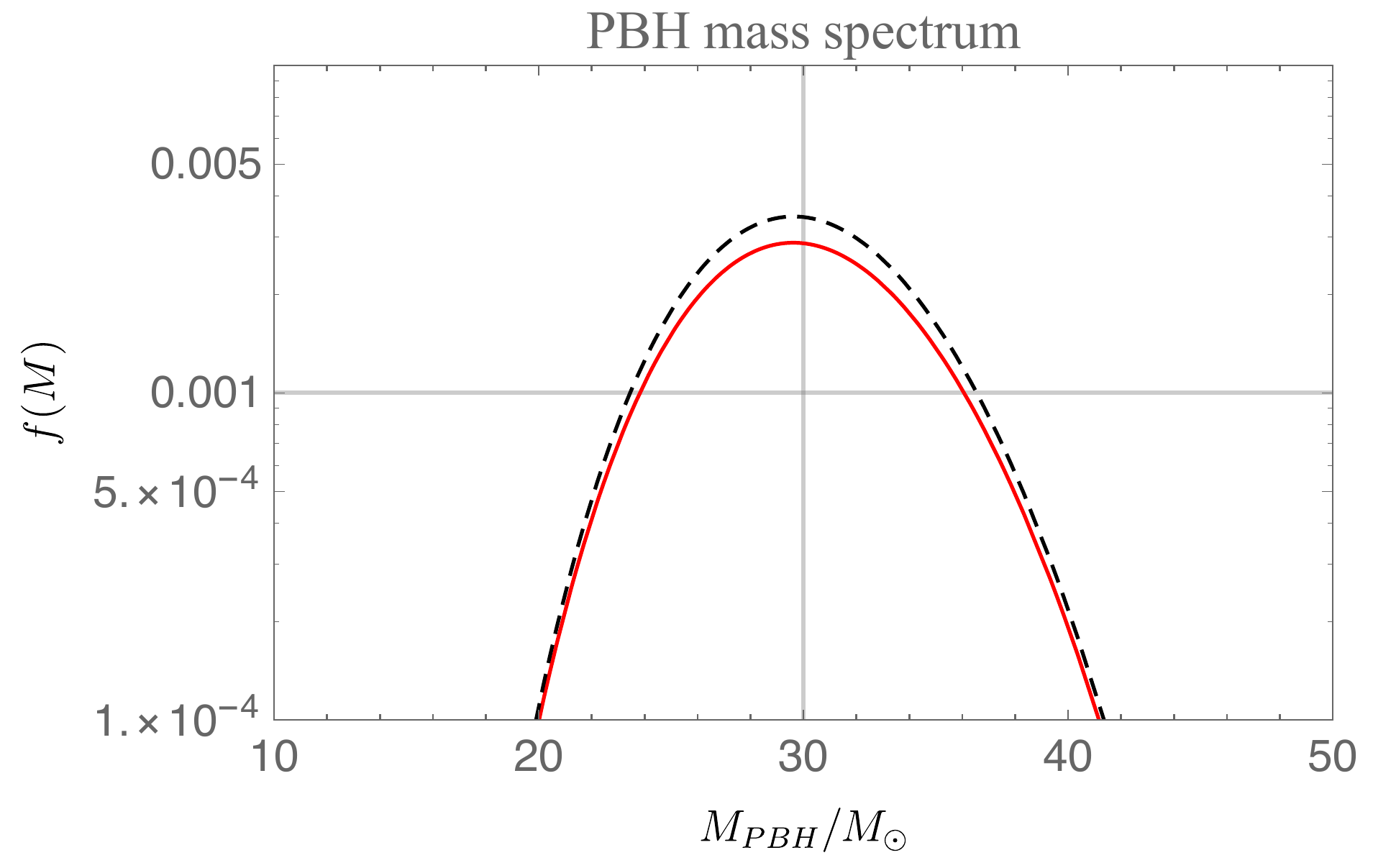}
			\end{center}\end{minipage}

		\begin{minipage}{0.55\hsize}	\begin{center}
			\includegraphics[width=1.\textwidth]{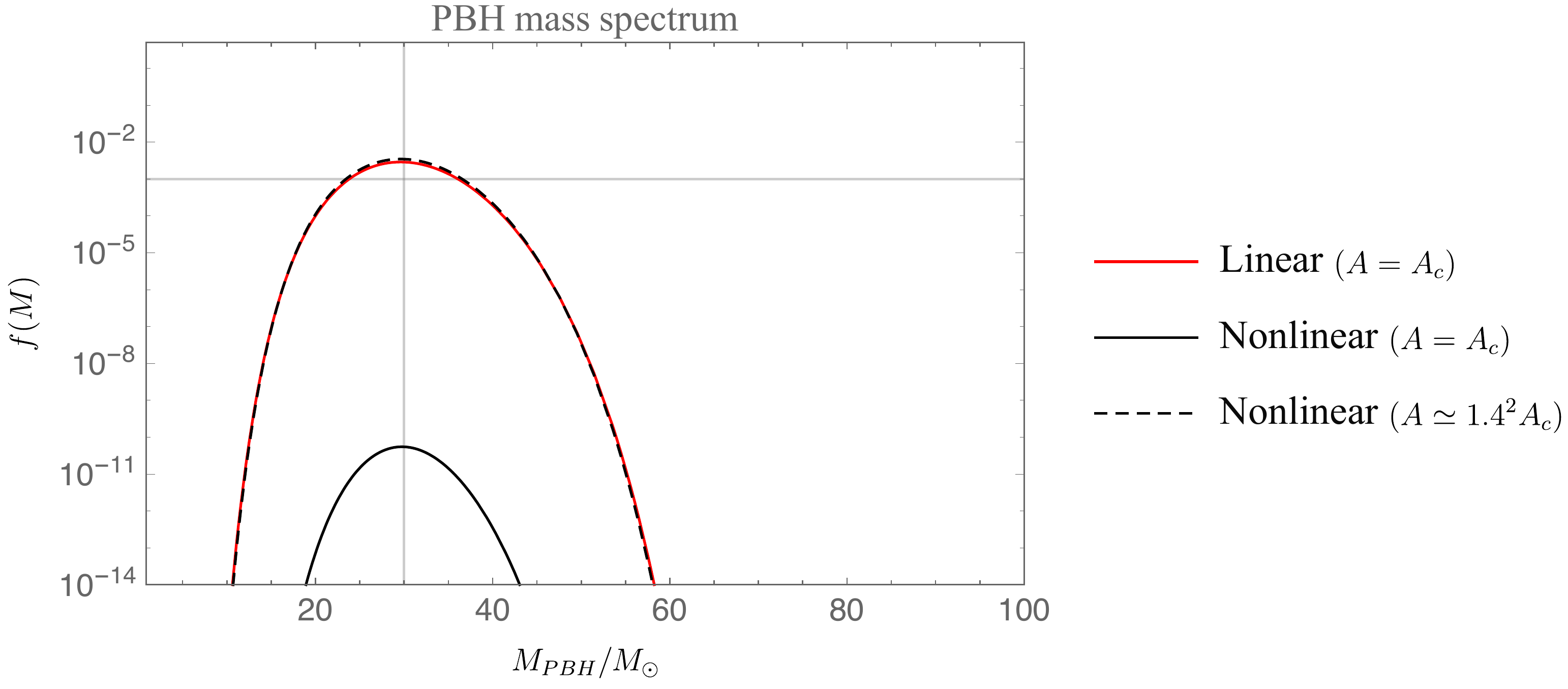}
		\end{center}  \end{minipage}
		
	\end{tabular}
	
	\caption{
		PBH mass spectrum on the linear and nonlinear calculations for the power spectrum \eqref{eq:powerspec}.
		We consider $30\msol$ PBHs for binary black holes observed in LIGO-Virgo collaboration.
		For linear order case in Eqs.\eqref{eq:gaussianP} and \eqref{eq:del2num}, we take the power spectrum with $k_*\sim 8.6 \times 10^{-5}\Mpc^{-1}$
		and $A_c=0.00605$, which satisfies $\Omega_\tx{PBH}/\Omega_\tx{DM}\simeq 10^{-3}$ .
		For nonlinear calculation in Eqs.\eqref{eq:del2num}, \eqref{eq:del3num} and \eqref{eq:probnongau3}, we take the two amplitude, $A_c$ and $A\simeq 1.4^2 A_c$, which comes from the analytical estimation in Sec.\ref{sec:NL}.  
	}
	\label{fig:massspct}
\end{figure}

Next, we calculate the mass spectrum of PBHs.
We consider $30\msol$ PBHs with  $\Omega_\tx{PBH}/\Omega_\tx{DM}\simeq 10^{-3}$ which account for the gravitational wave events discovered by the LIGO-Virgo collaboration \cite{Sasaki:2016jop} as a demonstration.
First, we consider the linear order calculation in Eq.\eqref{eq:gaussianP} with the top-hat window function and the transfer function, which is equivalent to Eq.\eqref{eq:probnongau3} with $\mu_3= 0$.
Inserting $\sigma$\eqref{eq:del2num} into the Gaussian probability distribution function~\eqref{eq:gaussianP}, we calculate the PBH mass spectrum \eqref{eq:massspec}.
We use the power spectrum \eqref{eq:powerspec} with $A_c=0.00605$ and $k_*\sim 8.6 \times 10^{5}\Mpc^{-1}$ to achieve $30\msol$ PBHs with $\Omega_\tx{PBH}/\Omega_\tx{DM}\simeq 10^{-3}$.
The red line in Fig.\ref{fig:massspct} shows the results of the linear order calculation.

Second, we include the skewness $\mu_3$ \eqref{eq:del3num} into the PBH formation rate \eqref{eq:probnongau3}.
The solid black line in Fig.\ref{fig:massspct} represent the mass spectrum on the nonlinear calculation on the power spectrum with $A=A_c$ and $k=k_*$.
As expected, the negative $\mu_3$ suppress the PBH formation rate and abundance.
Thus, the larger power spectrum is needed to reproduce the result of the linear calculation as shown by the black dashed line in Fig.~\ref{fig:massspct}.
It is seen that the nonlinear calculation well reproduce the linear one with the additional factor $\simeq 1.4^2$.

At last, we compare the non-Gaussianity from nonlinear relation between the density and curvature perturbations \eqref{eq:delHC} with the case where there would be primordial non-Gaussianity of the curvature perturbation.
When the curvature perturbation itself follows a non-Gaussian distribution, $\bar \delta_\tx{HC}$ also has non-Gaussianity.
As an example of non-Gaussianity of $\zeta$, we focus on the local form whose effect on the PBH formation was studied in~\cite{Young:2015cyn}.
The bispectrum $B_\zeta$ of the curvature perturbation is written as
\begin{align}
	\braket{\zeta(\bs k_1)\zeta(\bs k_2)\zeta(\bs k_3)}
	&=
	(2\pi)^3\delta^3(\bs k_1+\bs k_2+\bs k_3)
	B_\zeta(\bs k_1,\bs k_2,\bs k_3).
\\
	B_\zeta^\tx{local}(\bs k_1,\bs k_2,\bs k_3)
	&=
	\frac{6}{5}f_{NL}
	\left(
	P_\zeta(k_1)P_\zeta(k_2)
	+P_\zeta(k_2)P_\zeta(k_3)
	+P_\zeta(k_3)P_\zeta(k_1)
	\right),
\end{align}
where $f_{NL}$ is the non-linear parameter.
Using linear relation between $\zeta$ and $\bar\delta$ [Eqs.\eqref{eq:lindel} and \eqref{eq:deltaW}], we calculate the contribution of $B_\zeta^\tx{local}$ to the skewness of $\bar\delta$ as
\begin{align}
	\mu_{3:f_\tx{NL}}(r_m)
	&=\sigma^{-3}\braket{\bar \delta(\bs x, r_m)^3}
\nonumber
    \\	&=
    \sigma^{-3}
	\int \frac{\df^3 p}{(2\pi)^3}
	\frac{\df^3 q}{(2\pi)^3}
	\tilde W(p r_m)
	\tilde W(q r_m)
	\tilde W(|\bs p+\bs q| r_m)
\nonumber 
\\	&\quad\times \left( \frac{4}{9} \right)^3
	\left( pr_m qr_m |\bs p+\bs q|r_m \right)^2
	B_\zeta^{local}(\bs p,\bs q,-\bs p-\bs q)
\nonumber
\\	&=
    \sigma^{-3}
	\frac{18}{5}f_{NL}
	\left( \frac{4}{9} \right)^3
	\int  \frac{\df p}{p} 
	\tilde W(p r_m)\mathcal P_\zeta(p)
	(pr_m)^2
	\int \frac{\df q}{q} \tilde W(q r_m)\mathcal P_\zeta(q)
	(qr_m)^2
\nonumber\\&\quad\times
    \int^1_{-1}\frac{\df (\cos\theta)}{2}
	\tilde W(|\bs p+\bs q| r_m)
		\left(  |\bs p+\bs q|r_m \right)^2
\label{eq:delW3}
\end{align}
where $\cos\theta=\bs p\cdot\bs q/(|\bs p| |\bs q|)$ and $\sigma$ is obtained from \eqref{eq:oldsigma}.
Multiplying $\mathcal{P}_\zeta$ by the transfer function $T$, $\mu_{3:f_\tx{NL}}$ for the power spectrum \eqref{eq:powerspec} is given by
\begin{align}
	\mu_{3:f_\tx{NL}}(r_m)
	&=\sigma^{-3}
	A^2 \frac{18}{5}
	f_{NL}
	\left( \frac{4}{9} \right)^3
	|\tilde W(k_*r_m)|^2  |T(k_*r_m)|^4
	\left( k_*r_m \right)^4
	\\&\times
	\int^1_{-1}\frac{\df (\cos\theta)}{2}
	\tilde W(k_*r_m\sqrt{2+2\cos\theta} )
	\left( k_* r_m\sqrt{2+2\cos\theta} \right)^2,
	\label{eq:del3fnlnum}
\end{align}
We calculate $\mu_{3:f_\tx{NL}=1}(r_m)$ with the top-hat window function and show the result by the blue dashed line in Fig.\ref{fig:cumulants}. 
It is seen that the effect of the nonlinear $\bar\delta-\zeta$ relation is comparable to the primordial non-Gaussianity with $f_\tx{NL}\sim -1$.

\section{Conclusion}
\label{sec:cncl}

Using the long-wave approximation one obtains the density perturbation as a function of the curvature perturbation and it nonlinearly depends on the curvature perturbation.
We have studied the effect of the nonlinear relation on formation of PBHs.
As for the PBH formation criterion, we have adopted the compaction function which is also nonlinearly dependent on the curvature perturbation.
It is found that required curvature perturbation for the PBH formation increases due to the nonlinearity.

Since the criterion of PBH formation is obtained for spherically symmetric density profiles in the numerical simulation, we have used the spherical harmonics expansion of the curvature perturbation.
We then calculated the variance and the skewness of the density perturbation and derived the non-Gaussian property. 
We estimated the PBH abundance based on the Press–Schechter formalism with non-Gaussian probability density function.
As a result, we found that the nonlinear term slightly suppresses the PBH formation.
This suppression is comparable to that expected if the primordial curvature perturbation would have the local form of non-Gaussianity with non-linear parameter $f_\tx{NL}\sim -1$.
Comparing the linear order calculation, this suppression requires $\sim 1.4^2$ larger power spectrum.


\section*{Acknowledgements}
\small\noindent
This work was supported by JSPS KAKENHI Grant Nos. 17H01131 (M.K.) and 17K05434 (M.K.), MEXT KAKENHI Grant No. 15H05889 (M.K.), World Premier International Research Center Initiative (WPI Initiative), MEXT, Japan (M.K., H.N.), and Advanced Leading Graduate Course for Photon Science (H.N.).


\small
\bibliographystyle{apsrev4-1}
\bibliography{RefNC}

\begin{thebibliography}{38}%
\makeatletter
\providecommand \@ifxundefined [1]{%
 \@ifx{#1\undefined}
}%
\providecommand \@ifnum [1]{%
 \ifnum #1\expandafter \@firstoftwo
 \else \expandafter \@secondoftwo
 \fi
}%
\providecommand \@ifx [1]{%
 \ifx #1\expandafter \@firstoftwo
 \else \expandafter \@secondoftwo
 \fi
}%
\providecommand \natexlab [1]{#1}%
\providecommand \enquote  [1]{``#1''}%
\providecommand \bibnamefont  [1]{#1}%
\providecommand \bibfnamefont [1]{#1}%
\providecommand \citenamefont [1]{#1}%
\providecommand \href@noop [0]{\@secondoftwo}%
\providecommand \href [0]{\begingroup \@sanitize@url \@href}%
\providecommand \@href[1]{\@@startlink{#1}\@@href}%
\providecommand \@@href[1]{\endgroup#1\@@endlink}%
\providecommand \@sanitize@url [0]{\catcode `\\12\catcode `\$12\catcode
  `\&12\catcode `\#12\catcode `\^12\catcode `\_12\catcode `\%12\relax}%
\providecommand \@@startlink[1]{}%
\providecommand \@@endlink[0]{}%
\providecommand \url  [0]{\begingroup\@sanitize@url \@url }%
\providecommand \@url [1]{\endgroup\@href {#1}{\urlprefix }}%
\providecommand \urlprefix  [0]{URL }%
\providecommand \Eprint [0]{\href }%
\providecommand \doibase [0]{http://dx.doi.org/}%
\providecommand \selectlanguage [0]{\@gobble}%
\providecommand \bibinfo  [0]{\@secondoftwo}%
\providecommand \bibfield  [0]{\@secondoftwo}%
\providecommand \translation [1]{[#1]}%
\providecommand \BibitemOpen [0]{}%
\providecommand \bibitemStop [0]{}%
\providecommand \bibitemNoStop [0]{.\EOS\space}%
\providecommand \EOS [0]{\spacefactor3000\relax}%
\providecommand \BibitemShut  [1]{\csname bibitem#1\endcsname}%
\let\auto@bib@innerbib\@empty
\bibitem [{\citenamefont {Abbott}\ \emph {et~al.}(2018)\citenamefont {Abbott}
  \emph {et~al.}}]{LIGOScientific:2018jsj}%
  \BibitemOpen
  \bibfield  {author} {\bibinfo {author} {\bibfnamefont {B.~P.}\ \bibnamefont
  {Abbott}} \emph {et~al.} (\bibinfo {collaboration} {LIGO Scientific,
  Virgo}),\ }\href@noop {} {\  (\bibinfo {year} {2018})},\ \Eprint
  {http://arxiv.org/abs/1811.12940} {arXiv:1811.12940 [astro-ph.HE]}
  \BibitemShut {NoStop}%
\bibitem [{\citenamefont {Bird}\ \emph {et~al.}(2016)\citenamefont {Bird},
  \citenamefont {Cholis}, \citenamefont {Muñoz}, \citenamefont {Ali-Haïmoud},
  \citenamefont {Kamionkowski}, \citenamefont {Kovetz}, \citenamefont
  {Raccanelli},\ and\ \citenamefont {Riess}}]{Bird:2016dcv}%
  \BibitemOpen
  \bibfield  {author} {\bibinfo {author} {\bibfnamefont {S.}~\bibnamefont
  {Bird}}, \bibinfo {author} {\bibfnamefont {I.}~\bibnamefont {Cholis}},
  \bibinfo {author} {\bibfnamefont {J.~B.}\ \bibnamefont {Muñoz}}, \bibinfo
  {author} {\bibfnamefont {Y.}~\bibnamefont {Ali-Haïmoud}}, \bibinfo {author}
  {\bibfnamefont {M.}~\bibnamefont {Kamionkowski}}, \bibinfo {author}
  {\bibfnamefont {E.~D.}\ \bibnamefont {Kovetz}}, \bibinfo {author}
  {\bibfnamefont {A.}~\bibnamefont {Raccanelli}}, \ and\ \bibinfo {author}
  {\bibfnamefont {A.~G.}\ \bibnamefont {Riess}},\ }\href {\doibase
  10.1103/PhysRevLett.116.201301} {\bibfield  {journal} {\bibinfo  {journal}
  {Phys. Rev. Lett.}\ }\textbf {\bibinfo {volume} {116}},\ \bibinfo {pages}
  {201301} (\bibinfo {year} {2016})},\ \Eprint
  {http://arxiv.org/abs/1603.00464} {arXiv:1603.00464 [astro-ph.CO]}
  \BibitemShut {NoStop}%
\bibitem [{\citenamefont {Clesse}\ and\ \citenamefont
  {García-Bellido}(2017)}]{Clesse:2016vqa}%
  \BibitemOpen
  \bibfield  {author} {\bibinfo {author} {\bibfnamefont {S.}~\bibnamefont
  {Clesse}}\ and\ \bibinfo {author} {\bibfnamefont {J.}~\bibnamefont
  {García-Bellido}},\ }\href {\doibase 10.1016/j.dark.2016.10.002} {\bibfield
  {journal} {\bibinfo  {journal} {Phys. Dark Univ.}\ }\textbf {\bibinfo
  {volume} {15}},\ \bibinfo {pages} {142} (\bibinfo {year} {2017})},\ \Eprint
  {http://arxiv.org/abs/1603.05234} {arXiv:1603.05234 [astro-ph.CO]}
  \BibitemShut {NoStop}%
\bibitem [{\citenamefont {Sasaki}\ \emph {et~al.}(2016)\citenamefont {Sasaki},
  \citenamefont {Suyama}, \citenamefont {Tanaka},\ and\ \citenamefont
  {Yokoyama}}]{Sasaki:2016jop}%
  \BibitemOpen
  \bibfield  {author} {\bibinfo {author} {\bibfnamefont {M.}~\bibnamefont
  {Sasaki}}, \bibinfo {author} {\bibfnamefont {T.}~\bibnamefont {Suyama}},
  \bibinfo {author} {\bibfnamefont {T.}~\bibnamefont {Tanaka}}, \ and\ \bibinfo
  {author} {\bibfnamefont {S.}~\bibnamefont {Yokoyama}},\ }\href {\doibase
  10.1103/PhysRevLett.117.061101} {\bibfield  {journal} {\bibinfo  {journal}
  {Phys. Rev. Lett.}\ }\textbf {\bibinfo {volume} {117}},\ \bibinfo {pages}
  {061101} (\bibinfo {year} {2016})},\ \Eprint
  {http://arxiv.org/abs/1603.08338} {arXiv:1603.08338 [astro-ph.CO]}
  \BibitemShut {NoStop}%
\bibitem [{\citenamefont {Carr}\ \emph {et~al.}(2010)\citenamefont {Carr},
  \citenamefont {Kohri}, \citenamefont {Sendouda},\ and\ \citenamefont
  {Yokoyama}}]{Carr:2009jm}%
  \BibitemOpen
  \bibfield  {author} {\bibinfo {author} {\bibfnamefont {B.~J.}\ \bibnamefont
  {Carr}}, \bibinfo {author} {\bibfnamefont {K.}~\bibnamefont {Kohri}},
  \bibinfo {author} {\bibfnamefont {Y.}~\bibnamefont {Sendouda}}, \ and\
  \bibinfo {author} {\bibfnamefont {J.}~\bibnamefont {Yokoyama}},\ }\href
  {\doibase 10.1103/PhysRevD.81.104019} {\bibfield  {journal} {\bibinfo
  {journal} {Phys. Rev.}\ }\textbf {\bibinfo {volume} {D81}},\ \bibinfo {pages}
  {104019} (\bibinfo {year} {2010})},\ \Eprint {http://arxiv.org/abs/0912.5297}
  {arXiv:0912.5297 [astro-ph.CO]} \BibitemShut {NoStop}%
\bibitem [{\citenamefont {Allsman}\ \emph {et~al.}(2001)\citenamefont {Allsman}
  \emph {et~al.}}]{Allsman:2000kg}%
  \BibitemOpen
  \bibfield  {author} {\bibinfo {author} {\bibfnamefont {R.~A.}\ \bibnamefont
  {Allsman}} \emph {et~al.} (\bibinfo {collaboration} {Macho}),\ }\href
  {\doibase 10.1086/319636} {\bibfield  {journal} {\bibinfo  {journal}
  {Astrophys. J.}\ }\textbf {\bibinfo {volume} {550}},\ \bibinfo {pages} {L169}
  (\bibinfo {year} {2001})},\ \Eprint {http://arxiv.org/abs/astro-ph/0011506}
  {arXiv:astro-ph/0011506 [astro-ph]} \BibitemShut {NoStop}%
\bibitem [{\citenamefont {Tisserand}\ \emph {et~al.}(2007)\citenamefont
  {Tisserand} \emph {et~al.}}]{Tisserand:2006zx}%
  \BibitemOpen
  \bibfield  {author} {\bibinfo {author} {\bibfnamefont {P.}~\bibnamefont
  {Tisserand}} \emph {et~al.} (\bibinfo {collaboration} {EROS-2}),\ }\href
  {\doibase 10.1051/0004-6361:20066017} {\bibfield  {journal} {\bibinfo
  {journal} {Astron. Astrophys.}\ }\textbf {\bibinfo {volume} {469}},\ \bibinfo
  {pages} {387} (\bibinfo {year} {2007})},\ \Eprint
  {http://arxiv.org/abs/astro-ph/0607207} {arXiv:astro-ph/0607207 [astro-ph]}
  \BibitemShut {NoStop}%
\bibitem [{\citenamefont {Wyrzykowski}\ \emph {et~al.}(2011)\citenamefont
  {Wyrzykowski} \emph {et~al.}}]{Wyrzykowski:2011tr}%
  \BibitemOpen
  \bibfield  {author} {\bibinfo {author} {\bibfnamefont {L.}~\bibnamefont
  {Wyrzykowski}} \emph {et~al.},\ }\href {\doibase
  10.1111/j.1365-2966.2011.19243.x} {\bibfield  {journal} {\bibinfo  {journal}
  {Mon. Not. Roy. Astron. Soc.}\ }\textbf {\bibinfo {volume} {416}},\ \bibinfo
  {pages} {2949} (\bibinfo {year} {2011})},\ \Eprint
  {http://arxiv.org/abs/1106.2925} {arXiv:1106.2925 [astro-ph.GA]} \BibitemShut
  {NoStop}%
\bibitem [{\citenamefont {Niikura}\ \emph {et~al.}(2017)\citenamefont
  {Niikura}, \citenamefont {Takada}, \citenamefont {Yasuda}, \citenamefont
  {Lupton}, \citenamefont {Sumi}, \citenamefont {More}, \citenamefont {More},
  \citenamefont {Oguri},\ and\ \citenamefont {Chiba}}]{Niikura:2017zjd}%
  \BibitemOpen
  \bibfield  {author} {\bibinfo {author} {\bibfnamefont {H.}~\bibnamefont
  {Niikura}}, \bibinfo {author} {\bibfnamefont {M.}~\bibnamefont {Takada}},
  \bibinfo {author} {\bibfnamefont {N.}~\bibnamefont {Yasuda}}, \bibinfo
  {author} {\bibfnamefont {R.~H.}\ \bibnamefont {Lupton}}, \bibinfo {author}
  {\bibfnamefont {T.}~\bibnamefont {Sumi}}, \bibinfo {author} {\bibfnamefont
  {S.}~\bibnamefont {More}}, \bibinfo {author} {\bibfnamefont {A.}~\bibnamefont
  {More}}, \bibinfo {author} {\bibfnamefont {M.}~\bibnamefont {Oguri}}, \ and\
  \bibinfo {author} {\bibfnamefont {M.}~\bibnamefont {Chiba}},\ }\href@noop {}
  {\  (\bibinfo {year} {2017})},\ \Eprint {http://arxiv.org/abs/1701.02151}
  {arXiv:1701.02151 [astro-ph.CO]} \BibitemShut {NoStop}%
\bibitem [{\citenamefont {Ali-Haïmoud}\ and\ \citenamefont
  {Kamionkowski}(2017)}]{Ali-Haimoud:2016mbv}%
  \BibitemOpen
  \bibfield  {author} {\bibinfo {author} {\bibfnamefont {Y.}~\bibnamefont
  {Ali-Haïmoud}}\ and\ \bibinfo {author} {\bibfnamefont {M.}~\bibnamefont
  {Kamionkowski}},\ }\href {\doibase 10.1103/PhysRevD.95.043534} {\bibfield
  {journal} {\bibinfo  {journal} {Phys. Rev.}\ }\textbf {\bibinfo {volume}
  {D95}},\ \bibinfo {pages} {043534} (\bibinfo {year} {2017})},\ \Eprint
  {http://arxiv.org/abs/1612.05644} {arXiv:1612.05644 [astro-ph.CO]}
  \BibitemShut {NoStop}%
\bibitem [{\citenamefont {Graham}\ \emph {et~al.}(2015)\citenamefont {Graham},
  \citenamefont {Rajendran},\ and\ \citenamefont {Varela}}]{Graham:2015apa}%
  \BibitemOpen
  \bibfield  {author} {\bibinfo {author} {\bibfnamefont {P.~W.}\ \bibnamefont
  {Graham}}, \bibinfo {author} {\bibfnamefont {S.}~\bibnamefont {Rajendran}}, \
  and\ \bibinfo {author} {\bibfnamefont {J.}~\bibnamefont {Varela}},\ }\href
  {\doibase 10.1103/PhysRevD.92.063007} {\bibfield  {journal} {\bibinfo
  {journal} {Phys. Rev.}\ }\textbf {\bibinfo {volume} {D92}},\ \bibinfo {pages}
  {063007} (\bibinfo {year} {2015})},\ \Eprint
  {http://arxiv.org/abs/1505.04444} {arXiv:1505.04444 [hep-ph]} \BibitemShut
  {NoStop}%
\bibitem [{\citenamefont {Katz}\ \emph {et~al.}(2018)\citenamefont {Katz},
  \citenamefont {Kopp}, \citenamefont {Sibiryakov},\ and\ \citenamefont
  {Xue}}]{Katz:2018zrn}%
  \BibitemOpen
  \bibfield  {author} {\bibinfo {author} {\bibfnamefont {A.}~\bibnamefont
  {Katz}}, \bibinfo {author} {\bibfnamefont {J.}~\bibnamefont {Kopp}}, \bibinfo
  {author} {\bibfnamefont {S.}~\bibnamefont {Sibiryakov}}, \ and\ \bibinfo
  {author} {\bibfnamefont {W.}~\bibnamefont {Xue}},\ }\href {\doibase
  10.1088/1475-7516/2018/12/005} {\bibfield  {journal} {\bibinfo  {journal}
  {JCAP}\ }\textbf {\bibinfo {volume} {1812}},\ \bibinfo {pages} {005}
  (\bibinfo {year} {2018})},\ \Eprint {http://arxiv.org/abs/1807.11495}
  {arXiv:1807.11495 [astro-ph.CO]} \BibitemShut {NoStop}%
\bibitem [{\citenamefont {Bugaev}\ and\ \citenamefont
  {Klimai}(2012)}]{Bugaev:2011wy}%
  \BibitemOpen
  \bibfield  {author} {\bibinfo {author} {\bibfnamefont {E.}~\bibnamefont
  {Bugaev}}\ and\ \bibinfo {author} {\bibfnamefont {P.}~\bibnamefont
  {Klimai}},\ }\href {\doibase 10.1103/PhysRevD.85.103504} {\bibfield
  {journal} {\bibinfo  {journal} {Phys. Rev.}\ }\textbf {\bibinfo {volume}
  {D85}},\ \bibinfo {pages} {103504} (\bibinfo {year} {2012})},\ \Eprint
  {http://arxiv.org/abs/1112.5601} {arXiv:1112.5601 [astro-ph.CO]} \BibitemShut
  {NoStop}%
\bibitem [{\citenamefont {Inomata}\ \emph {et~al.}(2018)\citenamefont
  {Inomata}, \citenamefont {Kawasaki}, \citenamefont {Mukaida},\ and\
  \citenamefont {Yanagida}}]{Inomata:2017vxo}%
  \BibitemOpen
  \bibfield  {author} {\bibinfo {author} {\bibfnamefont {K.}~\bibnamefont
  {Inomata}}, \bibinfo {author} {\bibfnamefont {M.}~\bibnamefont {Kawasaki}},
  \bibinfo {author} {\bibfnamefont {K.}~\bibnamefont {Mukaida}}, \ and\
  \bibinfo {author} {\bibfnamefont {T.~T.}\ \bibnamefont {Yanagida}},\ }\href
  {\doibase 10.1103/PhysRevD.97.043514} {\bibfield  {journal} {\bibinfo
  {journal} {Phys. Rev.}\ }\textbf {\bibinfo {volume} {D97}},\ \bibinfo {pages}
  {043514} (\bibinfo {year} {2018})},\ \Eprint
  {http://arxiv.org/abs/1711.06129} {arXiv:1711.06129 [astro-ph.CO]}
  \BibitemShut {NoStop}%
\bibitem [{\citenamefont {Kohri}\ \emph {et~al.}(2013)\citenamefont {Kohri},
  \citenamefont {Lin},\ and\ \citenamefont {Matsuda}}]{Kohri:2012yw}%
  \BibitemOpen
  \bibfield  {author} {\bibinfo {author} {\bibfnamefont {K.}~\bibnamefont
  {Kohri}}, \bibinfo {author} {\bibfnamefont {C.-M.}\ \bibnamefont {Lin}}, \
  and\ \bibinfo {author} {\bibfnamefont {T.}~\bibnamefont {Matsuda}},\ }\href
  {\doibase 10.1103/PhysRevD.87.103527} {\bibfield  {journal} {\bibinfo
  {journal} {Phys. Rev.}\ }\textbf {\bibinfo {volume} {D87}},\ \bibinfo {pages}
  {103527} (\bibinfo {year} {2013})},\ \Eprint {http://arxiv.org/abs/1211.2371}
  {arXiv:1211.2371 [hep-ph]} \BibitemShut {NoStop}%
\bibitem [{\citenamefont {Ando}\ \emph {et~al.}(2017)\citenamefont {Ando},
  \citenamefont {Inomata}, \citenamefont {Kawasaki}, \citenamefont {Mukaida},\
  and\ \citenamefont {Yanagida}}]{Ando:2017veq}%
  \BibitemOpen
  \bibfield  {author} {\bibinfo {author} {\bibfnamefont {K.}~\bibnamefont
  {Ando}}, \bibinfo {author} {\bibfnamefont {K.}~\bibnamefont {Inomata}},
  \bibinfo {author} {\bibfnamefont {M.}~\bibnamefont {Kawasaki}}, \bibinfo
  {author} {\bibfnamefont {K.}~\bibnamefont {Mukaida}}, \ and\ \bibinfo
  {author} {\bibfnamefont {T.~T.}\ \bibnamefont {Yanagida}},\ }\href@noop {} {\
   (\bibinfo {year} {2017})},\ \Eprint {http://arxiv.org/abs/1711.08956}
  {arXiv:1711.08956 [astro-ph.CO]} \BibitemShut {NoStop}%
\bibitem [{\citenamefont {Carr}(1975)}]{Carr:1975qj}%
  \BibitemOpen
  \bibfield  {author} {\bibinfo {author} {\bibfnamefont {B.~J.}\ \bibnamefont
  {Carr}},\ }\href {\doibase 10.1086/153853} {\bibfield  {journal} {\bibinfo
  {journal} {Astrophys. J.}\ }\textbf {\bibinfo {volume} {201}},\ \bibinfo
  {pages} {1} (\bibinfo {year} {1975})}\BibitemShut {NoStop}%
\bibitem [{\citenamefont {Shibata}\ and\ \citenamefont
  {Sasaki}(1999)}]{Shibata:1999zs}%
  \BibitemOpen
  \bibfield  {author} {\bibinfo {author} {\bibfnamefont {M.}~\bibnamefont
  {Shibata}}\ and\ \bibinfo {author} {\bibfnamefont {M.}~\bibnamefont
  {Sasaki}},\ }\href {\doibase 10.1103/PhysRevD.60.084002} {\bibfield
  {journal} {\bibinfo  {journal} {Phys. Rev.}\ }\textbf {\bibinfo {volume}
  {D60}},\ \bibinfo {pages} {084002} (\bibinfo {year} {1999})},\ \Eprint
  {http://arxiv.org/abs/gr-qc/9905064} {arXiv:gr-qc/9905064 [gr-qc]}
  \BibitemShut {NoStop}%
\bibitem [{\citenamefont {Polnarev}\ and\ \citenamefont
  {Musco}(2007)}]{Polnarev:2006aa}%
  \BibitemOpen
  \bibfield  {author} {\bibinfo {author} {\bibfnamefont {A.~G.}\ \bibnamefont
  {Polnarev}}\ and\ \bibinfo {author} {\bibfnamefont {I.}~\bibnamefont
  {Musco}},\ }\href {\doibase 10.1088/0264-9381/24/6/003} {\bibfield  {journal}
  {\bibinfo  {journal} {Class. Quant. Grav.}\ }\textbf {\bibinfo {volume}
  {24}},\ \bibinfo {pages} {1405} (\bibinfo {year} {2007})},\ \Eprint
  {http://arxiv.org/abs/gr-qc/0605122} {arXiv:gr-qc/0605122 [gr-qc]}
  \BibitemShut {NoStop}%
\bibitem [{\citenamefont {Harada}\ \emph {et~al.}(2015)\citenamefont {Harada},
  \citenamefont {Yoo}, \citenamefont {Nakama},\ and\ \citenamefont
  {Koga}}]{Harada:2015yda}%
  \BibitemOpen
  \bibfield  {author} {\bibinfo {author} {\bibfnamefont {T.}~\bibnamefont
  {Harada}}, \bibinfo {author} {\bibfnamefont {C.-M.}\ \bibnamefont {Yoo}},
  \bibinfo {author} {\bibfnamefont {T.}~\bibnamefont {Nakama}}, \ and\ \bibinfo
  {author} {\bibfnamefont {Y.}~\bibnamefont {Koga}},\ }\href {\doibase
  10.1103/PhysRevD.91.084057} {\bibfield  {journal} {\bibinfo  {journal} {Phys.
  Rev.}\ }\textbf {\bibinfo {volume} {D91}},\ \bibinfo {pages} {084057}
  (\bibinfo {year} {2015})},\ \Eprint {http://arxiv.org/abs/1503.03934}
  {arXiv:1503.03934 [gr-qc]} \BibitemShut {NoStop}%
\bibitem [{\citenamefont {Young}\ \emph {et~al.}(2014)\citenamefont {Young},
  \citenamefont {Byrnes},\ and\ \citenamefont {Sasaki}}]{Young:2014ana}%
  \BibitemOpen
  \bibfield  {author} {\bibinfo {author} {\bibfnamefont {S.}~\bibnamefont
  {Young}}, \bibinfo {author} {\bibfnamefont {C.~T.}\ \bibnamefont {Byrnes}}, \
  and\ \bibinfo {author} {\bibfnamefont {M.}~\bibnamefont {Sasaki}},\ }\href
  {\doibase 10.1088/1475-7516/2014/07/045} {\bibfield  {journal} {\bibinfo
  {journal} {JCAP}\ }\textbf {\bibinfo {volume} {1407}},\ \bibinfo {pages}
  {045} (\bibinfo {year} {2014})},\ \Eprint {http://arxiv.org/abs/1405.7023}
  {arXiv:1405.7023 [gr-qc]} \BibitemShut {NoStop}%
\bibitem [{\citenamefont {Press}\ and\ \citenamefont
  {Schechter}(1974)}]{Press:1973iz}%
  \BibitemOpen
  \bibfield  {author} {\bibinfo {author} {\bibfnamefont {W.~H.}\ \bibnamefont
  {Press}}\ and\ \bibinfo {author} {\bibfnamefont {P.}~\bibnamefont
  {Schechter}},\ }\href {\doibase 10.1086/152650} {\bibfield  {journal}
  {\bibinfo  {journal} {Astrophys. J.}\ }\textbf {\bibinfo {volume} {187}},\
  \bibinfo {pages} {425} (\bibinfo {year} {1974})}\BibitemShut {NoStop}%
\bibitem [{\citenamefont {Bardeen}\ \emph {et~al.}(1986)\citenamefont
  {Bardeen}, \citenamefont {Bond}, \citenamefont {Kaiser},\ and\ \citenamefont
  {Szalay}}]{Bardeen:1985tr}%
  \BibitemOpen
  \bibfield  {author} {\bibinfo {author} {\bibfnamefont {J.~M.}\ \bibnamefont
  {Bardeen}}, \bibinfo {author} {\bibfnamefont {J.~R.}\ \bibnamefont {Bond}},
  \bibinfo {author} {\bibfnamefont {N.}~\bibnamefont {Kaiser}}, \ and\ \bibinfo
  {author} {\bibfnamefont {A.~S.}\ \bibnamefont {Szalay}},\ }\href {\doibase
  10.1086/164143} {\bibfield  {journal} {\bibinfo  {journal} {Astrophys. J.}\
  }\textbf {\bibinfo {volume} {304}},\ \bibinfo {pages} {15} (\bibinfo {year}
  {1986})}\BibitemShut {NoStop}%
\bibitem [{\citenamefont {Green}\ \emph {et~al.}(2004)\citenamefont {Green},
  \citenamefont {Liddle}, \citenamefont {Malik},\ and\ \citenamefont
  {Sasaki}}]{Green:2004wb}%
  \BibitemOpen
  \bibfield  {author} {\bibinfo {author} {\bibfnamefont {A.~M.}\ \bibnamefont
  {Green}}, \bibinfo {author} {\bibfnamefont {A.~R.}\ \bibnamefont {Liddle}},
  \bibinfo {author} {\bibfnamefont {K.~A.}\ \bibnamefont {Malik}}, \ and\
  \bibinfo {author} {\bibfnamefont {M.}~\bibnamefont {Sasaki}},\ }\href
  {\doibase 10.1103/PhysRevD.70.041502} {\bibfield  {journal} {\bibinfo
  {journal} {Phys. Rev.}\ }\textbf {\bibinfo {volume} {D70}},\ \bibinfo {pages}
  {041502} (\bibinfo {year} {2004})},\ \Eprint
  {http://arxiv.org/abs/astro-ph/0403181} {arXiv:astro-ph/0403181 [astro-ph]}
  \BibitemShut {NoStop}%
\bibitem [{\citenamefont {Yoo}\ \emph {et~al.}(2018)\citenamefont {Yoo},
  \citenamefont {Harada}, \citenamefont {Garriga},\ and\ \citenamefont
  {Kohri}}]{Yoo:2018kvb}%
  \BibitemOpen
  \bibfield  {author} {\bibinfo {author} {\bibfnamefont {C.-M.}\ \bibnamefont
  {Yoo}}, \bibinfo {author} {\bibfnamefont {T.}~\bibnamefont {Harada}},
  \bibinfo {author} {\bibfnamefont {J.}~\bibnamefont {Garriga}}, \ and\
  \bibinfo {author} {\bibfnamefont {K.}~\bibnamefont {Kohri}},\ }\href@noop {}
  {\  (\bibinfo {year} {2018})},\ \Eprint {http://arxiv.org/abs/1805.03946}
  {arXiv:1805.03946 [astro-ph.CO]} \BibitemShut {NoStop}%
\bibitem [{\citenamefont {Maldacena}(2003)}]{Maldacena:2002vr}%
  \BibitemOpen
  \bibfield  {author} {\bibinfo {author} {\bibfnamefont {J.~M.}\ \bibnamefont
  {Maldacena}},\ }\href {\doibase 10.1088/1126-6708/2003/05/013} {\bibfield
  {journal} {\bibinfo  {journal} {JHEP}\ }\textbf {\bibinfo {volume} {05}},\
  \bibinfo {pages} {013} (\bibinfo {year} {2003})},\ \Eprint
  {http://arxiv.org/abs/astro-ph/0210603} {arXiv:astro-ph/0210603 [astro-ph]}
  \BibitemShut {NoStop}%
\bibitem [{\citenamefont {Pina~Avelino}(2005)}]{PinaAvelino:2005rm}%
  \BibitemOpen
  \bibfield  {author} {\bibinfo {author} {\bibfnamefont {P.}~\bibnamefont
  {Pina~Avelino}},\ }\href {\doibase 10.1103/PhysRevD.72.124004} {\bibfield
  {journal} {\bibinfo  {journal} {Phys. Rev.}\ }\textbf {\bibinfo {volume}
  {D72}},\ \bibinfo {pages} {124004} (\bibinfo {year} {2005})},\ \Eprint
  {http://arxiv.org/abs/astro-ph/0510052} {arXiv:astro-ph/0510052 [astro-ph]}
  \BibitemShut {NoStop}%
\bibitem [{\citenamefont {Byrnes}\ \emph {et~al.}(2012)\citenamefont {Byrnes},
  \citenamefont {Copeland},\ and\ \citenamefont {Green}}]{Byrnes:2012yx}%
  \BibitemOpen
  \bibfield  {author} {\bibinfo {author} {\bibfnamefont {C.~T.}\ \bibnamefont
  {Byrnes}}, \bibinfo {author} {\bibfnamefont {E.~J.}\ \bibnamefont
  {Copeland}}, \ and\ \bibinfo {author} {\bibfnamefont {A.~M.}\ \bibnamefont
  {Green}},\ }\href {\doibase 10.1103/PhysRevD.86.043512} {\bibfield  {journal}
  {\bibinfo  {journal} {Phys. Rev.}\ }\textbf {\bibinfo {volume} {D86}},\
  \bibinfo {pages} {043512} (\bibinfo {year} {2012})},\ \Eprint
  {http://arxiv.org/abs/1206.4188} {arXiv:1206.4188 [astro-ph.CO]} \BibitemShut
  {NoStop}%
\bibitem [{\citenamefont {Young}\ and\ \citenamefont
  {Byrnes}(2013)}]{Young:2013oia}%
  \BibitemOpen
  \bibfield  {author} {\bibinfo {author} {\bibfnamefont {S.}~\bibnamefont
  {Young}}\ and\ \bibinfo {author} {\bibfnamefont {C.~T.}\ \bibnamefont
  {Byrnes}},\ }\href {\doibase 10.1088/1475-7516/2013/08/052} {\bibfield
  {journal} {\bibinfo  {journal} {JCAP}\ }\textbf {\bibinfo {volume} {1308}},\
  \bibinfo {pages} {052} (\bibinfo {year} {2013})},\ \Eprint
  {http://arxiv.org/abs/1307.4995} {arXiv:1307.4995 [astro-ph.CO]} \BibitemShut
  {NoStop}%
\bibitem [{\citenamefont {Sherkatghanad}\ and\ \citenamefont
  {Brandenberger}(2015)}]{Sherkatghanad:2015rga}%
  \BibitemOpen
  \bibfield  {author} {\bibinfo {author} {\bibfnamefont {Z.}~\bibnamefont
  {Sherkatghanad}}\ and\ \bibinfo {author} {\bibfnamefont {R.~H.}\ \bibnamefont
  {Brandenberger}},\ }\href@noop {} {\  (\bibinfo {year} {2015})},\ \Eprint
  {http://arxiv.org/abs/1508.00968} {arXiv:1508.00968 [astro-ph.CO]}
  \BibitemShut {NoStop}%
\bibitem [{\citenamefont {Young}\ \emph {et~al.}(2016)\citenamefont {Young},
  \citenamefont {Regan},\ and\ \citenamefont {Byrnes}}]{Young:2015cyn}%
  \BibitemOpen
  \bibfield  {author} {\bibinfo {author} {\bibfnamefont {S.}~\bibnamefont
  {Young}}, \bibinfo {author} {\bibfnamefont {D.}~\bibnamefont {Regan}}, \ and\
  \bibinfo {author} {\bibfnamefont {C.~T.}\ \bibnamefont {Byrnes}},\ }\href
  {\doibase 10.1088/1475-7516/2016/02/029} {\bibfield  {journal} {\bibinfo
  {journal} {JCAP}\ }\textbf {\bibinfo {volume} {1602}},\ \bibinfo {pages}
  {029} (\bibinfo {year} {2016})},\ \Eprint {http://arxiv.org/abs/1512.07224}
  {arXiv:1512.07224 [astro-ph.CO]} \BibitemShut {NoStop}%
\bibitem [{\citenamefont {Franciolini}\ \emph {et~al.}(2018)\citenamefont
  {Franciolini}, \citenamefont {Kehagias}, \citenamefont {Matarrese},\ and\
  \citenamefont {Riotto}}]{Franciolini:2018vbk}%
  \BibitemOpen
  \bibfield  {author} {\bibinfo {author} {\bibfnamefont {G.}~\bibnamefont
  {Franciolini}}, \bibinfo {author} {\bibfnamefont {A.}~\bibnamefont
  {Kehagias}}, \bibinfo {author} {\bibfnamefont {S.}~\bibnamefont {Matarrese}},
  \ and\ \bibinfo {author} {\bibfnamefont {A.}~\bibnamefont {Riotto}},\ }\href
  {\doibase 10.1088/1475-7516/2018/03/016} {\bibfield  {journal} {\bibinfo
  {journal} {JCAP}\ }\textbf {\bibinfo {volume} {1803}},\ \bibinfo {pages}
  {016} (\bibinfo {year} {2018})},\ \Eprint {http://arxiv.org/abs/1801.09415}
  {arXiv:1801.09415 [astro-ph.CO]} \BibitemShut {NoStop}%
\bibitem [{\citenamefont {Harada}\ \emph {et~al.}(2013)\citenamefont {Harada},
  \citenamefont {Yoo},\ and\ \citenamefont {Kohri}}]{Harada:2013epa}%
  \BibitemOpen
  \bibfield  {author} {\bibinfo {author} {\bibfnamefont {T.}~\bibnamefont
  {Harada}}, \bibinfo {author} {\bibfnamefont {C.-M.}\ \bibnamefont {Yoo}}, \
  and\ \bibinfo {author} {\bibfnamefont {K.}~\bibnamefont {Kohri}},\ }\href
  {\doibase 10.1103/PhysRevD.88.084051, 10.1103/PhysRevD.89.029903} {\bibfield
  {journal} {\bibinfo  {journal} {Phys. Rev.}\ }\textbf {\bibinfo {volume}
  {D88}},\ \bibinfo {pages} {084051} (\bibinfo {year} {2013})},\ \bibinfo
  {note} {[Erratum: Phys. Rev.D89,no.2,029903(2014)]},\ \Eprint
  {http://arxiv.org/abs/1309.4201} {arXiv:1309.4201 [astro-ph.CO]} \BibitemShut
  {NoStop}%
\bibitem [{\citenamefont {Nakama}\ \emph {et~al.}(2014)\citenamefont {Nakama},
  \citenamefont {Harada}, \citenamefont {Polnarev},\ and\ \citenamefont
  {Yokoyama}}]{Nakama:2013ica}%
  \BibitemOpen
  \bibfield  {author} {\bibinfo {author} {\bibfnamefont {T.}~\bibnamefont
  {Nakama}}, \bibinfo {author} {\bibfnamefont {T.}~\bibnamefont {Harada}},
  \bibinfo {author} {\bibfnamefont {A.~G.}\ \bibnamefont {Polnarev}}, \ and\
  \bibinfo {author} {\bibfnamefont {J.}~\bibnamefont {Yokoyama}},\ }\href
  {\doibase 10.1088/1475-7516/2014/01/037} {\bibfield  {journal} {\bibinfo
  {journal} {JCAP}\ }\textbf {\bibinfo {volume} {1401}},\ \bibinfo {pages}
  {037} (\bibinfo {year} {2014})},\ \Eprint {http://arxiv.org/abs/1310.3007}
  {arXiv:1310.3007 [gr-qc]} \BibitemShut {NoStop}%
\bibitem [{\citenamefont {Ando}\ \emph {et~al.}(2018)\citenamefont {Ando},
  \citenamefont {Inomata},\ and\ \citenamefont {Kawasaki}}]{Ando:2018qdb}%
  \BibitemOpen
  \bibfield  {author} {\bibinfo {author} {\bibfnamefont {K.}~\bibnamefont
  {Ando}}, \bibinfo {author} {\bibfnamefont {K.}~\bibnamefont {Inomata}}, \
  and\ \bibinfo {author} {\bibfnamefont {M.}~\bibnamefont {Kawasaki}},\
  }\href@noop {} {\  (\bibinfo {year} {2018})},\ \Eprint
  {http://arxiv.org/abs/1802.06393} {arXiv:1802.06393 [astro-ph.CO]}
  \BibitemShut {NoStop}%
\bibitem [{\citenamefont {Yoo}\ and\ \citenamefont
  {Desjacques}(2013)}]{Yoo:2013tc}%
  \BibitemOpen
  \bibfield  {author} {\bibinfo {author} {\bibfnamefont {J.}~\bibnamefont
  {Yoo}}\ and\ \bibinfo {author} {\bibfnamefont {V.}~\bibnamefont
  {Desjacques}},\ }\href {\doibase 10.1103/PhysRevD.88.023502} {\bibfield
  {journal} {\bibinfo  {journal} {Phys. Rev.}\ }\textbf {\bibinfo {volume}
  {D88}},\ \bibinfo {pages} {023502} (\bibinfo {year} {2013})},\ \Eprint
  {http://arxiv.org/abs/1301.4501} {arXiv:1301.4501 [astro-ph.CO]} \BibitemShut
  {NoStop}%
\bibitem [{\citenamefont {Ezquiaga}\ and\ \citenamefont
  {García-Bellido}(2018)}]{Ezquiaga:2018gbw}%
  \BibitemOpen
  \bibfield  {author} {\bibinfo {author} {\bibfnamefont {J.~M.}\ \bibnamefont
  {Ezquiaga}}\ and\ \bibinfo {author} {\bibfnamefont {J.}~\bibnamefont
  {García-Bellido}},\ }\href@noop {} {\  (\bibinfo {year} {2018})},\ \Eprint
  {http://arxiv.org/abs/1805.06731} {arXiv:1805.06731 [astro-ph.CO]}
  \BibitemShut {NoStop}%
\bibitem [{\citenamefont {Niemeyer}\ and\ \citenamefont
  {Jedamzik}(1999)}]{Niemeyer:1999ak}%
  \BibitemOpen
  \bibfield  {author} {\bibinfo {author} {\bibfnamefont {J.~C.}\ \bibnamefont
  {Niemeyer}}\ and\ \bibinfo {author} {\bibfnamefont {K.}~\bibnamefont
  {Jedamzik}},\ }\href {\doibase 10.1103/PhysRevD.59.124013} {\bibfield
  {journal} {\bibinfo  {journal} {Phys. Rev.}\ }\textbf {\bibinfo {volume}
  {D59}},\ \bibinfo {pages} {124013} (\bibinfo {year} {1999})},\ \Eprint
  {http://arxiv.org/abs/astro-ph/9901292} {arXiv:astro-ph/9901292 [astro-ph]}
  \BibitemShut {NoStop}%
\end{thebibliography}%
\end{document}